\documentclass[aps,prb,twocolumn,superscriptaddress,showpacs]{revtex4}
\usepackage{graphicx}
\usepackage{epsfig}
\usepackage{latexsym}
\usepackage[dvips]{color}

\begin{document}


\title{From Cooperative Paramagnet to N\'{e}el Order in Y$_2$Ru$_2$O$_7$}


\author{J. van Duijn}
\affiliation{Department of Physics and Astronomy, The Johns Hopkins University, Baltimore, Maryland 21218}
\affiliation{ISIS Facility, Rutherford Appleton Laboratory, Chilton, Didcot, OX11 0QX, U.K.}
\affiliation{Dept. Quim. Inorgan., Fac. Quim., Unversidad Complutense de Madrid, Madrid, E-28040, Spain.}
\author{N. Hur}
\affiliation{Rutgers Center for Emergent Materials and Department of Physics and Astronomy, Rutgers University, Piscataway, New Jersey 08854}
\author{J. W. Taylor}
\affiliation{ISIS Facility, Rutherford Appleton Laboratory, Chilton, Didcot, OX11 0QX, U.K.}
\author{Y. Qiu}
\affiliation{NIST Center for Neutron Research, National Institute of Standards and Technology, Gaithersburg, Maryland 20899}
\author{Q. Z. Huang}
\affiliation{NIST Center for Neutron Research, National Institute of Standards and Technology, Gaithersburg, Maryland 20899}
\author{S.-W. Cheong}
\affiliation{Rutgers Center for Emergent Materials and Department of Physics and Astronomy, Rutgers University, Piscataway, New Jersey 08854}
\author{C. Broholm}
\affiliation{Department of Physics and Astronomy, The Johns Hopkins University, Baltimore, Maryland 21218}
\affiliation{NIST Center for Neutron Research, National Institute of Standards and Technology, Gaithersburg, Maryland 20899}
\author{T. G. Perring}
\affiliation{ISIS Facility, Rutherford Appleton Laboratory, Chilton, Didcot, OX11 0QX, U.K.}
\date{\today}

\begin{abstract}
Spin correlations in the pyrochlore antiferromagnet Y$_2$Ru$_2$O$_7$ with Curie-Weiss temperature $\Theta_{CW}=-1100$~K and critical temperature $T_N=77$~K were examined through neutron scattering. For $T_N<T<\Theta_{CW}/3$ the data show spin relaxation with a rate $\hbar\Gamma= 1.17(9)k_BT$. For $T<T_N$ spectral weight moves to higher energies with substantial changes up to $4\times k_BT_N$. For $T<<T_N$ there is a $\Delta=11(1)$~meV energy gap and a pronounced spectral maximum at 19.7 meV. Throughout the temperature range examined the wave vector dependence of inelastic scattering exhibits a broad peak for $Qd\approx 3.8$ ($d$ is Ru-Ru spacing) consistent with dipolar spin correlations.
\end{abstract}

\pacs{76.50.+g, 75.40.Gb, 75.50.Ee}

\maketitle

In the pyrochlore structure, with general formula $\rm A_2B_2O_7$, both the A (trivalent rare earth ions) and B (tetravalent transition metal ions) site ions form a lattice of corner sharing tetrahedra.\cite{frustration} Weakly connected and non-bipartite, this lattice produces anomalous frustrated magnetism when populated by spins with near neighbor antiferromagnetic (AFM) interactions.\cite{anderson} For classical spin vectors the lowest energy states form a manifold characterized by zero magnetization on every tetrahedron.~\cite{chalkermoessner} Rather than N\'{e}el order for $T\approx |\Theta_{CW}|$, pyrochlore magnets remain disordered to much lower $T$ where the collective properties are governed by quantum fluctuations, thermal fluctuations, and sub-leading interactions.~\cite{isakov,henley,tchernyshyovpyro} Examples of low $T$ states that have been reported are spin glasses
\cite{pyro_1}, spin liquids\cite{tb2ti2o7,pr2ir2o7}, and magneto-elastically induced N\'{e}el order.~\cite{ZnCr2O4prl,spinp}

Y$_2$Ru$_2$O$_7$ is a Mott insulator with $S= 1$ Ru$^{4+}$ spin and a large negative Curie-Weiss temperature $\Theta_{CW}\approx$ -1100 K.~\cite{MI_1,MI_2, MI_3,CW} Specific heat and susceptibility indicate a magnetic phase transition at $T_N$= 77 K.~\cite{MI_2, Cp, spin_gap} Diffraction measurements show that the transition is to a long range ordered (LRO) AFM $q= 0$ structure.~\cite{diffraction} The total spin per tetrahedron vanishes so this state is indeed a specific member of the low energy AFM pyrochlore manifold. Here we report a neutron scattering experiment probing the anomalous spin dynamics of Y$_2$Ru$_2$O$_7$. The measurements reveal that AFM spin correlations are established at temperatures at least as high as 300 K and evolve little upon cooling. The spin relaxation rate however decreases linearly in $T$ according to $\hbar\Gamma= 1.17k_BT$ consistent with proximity to quantum criticality. However N\'{e}el order yields an 11(1) meV gap in the excitation spectrum and moves spectral weight to a 19.7 meV peak.

A powder sample of Y$_2$Ru$_2$O$_7$ was synthesised using the solid state reaction method. A mixture of Y$_2$O$_3$ and RuO$_2$ in proper molar ratio was pre-reacted at 850 $^\circ$C for 15 h in air and then ground and pressed to pellets. The pellets were sintered at 1000-1200 $^\circ$C in air with intermediate grindings to improve homogeneity. Powder neutron diffraction data were collected using the BT1 diffractometer at NIST. Rietveld analysis confirmed the cubic pyrochlore structure above and below $T_N$ with spacegroup Fd$\overline{3}$m (227) and lattice parameter $a=10.12824(2)$~\AA\ at $T=10$~K. The analysis also revealed 0.044(2) weight \%  unreacted Y$_2$O$_3$. Inelastic neutron scattering experiments were performed on the MARI spectrometer at the ISIS Facility and the DCS spectrometer at NIST. On MARI we used an incident energy E$_i$= 150 meV for an elastic energy resolution of 3.5 meV. The phonon contribution to neutron scattering was subtracted by scaling the spectrum measured at high wave vectors, where the magnetic response is negligible, using the Discus package.~\cite{discus} This program uses a Monte Carlo method to calculate a wave vector dependent scaling factor, which showed no significant energy dependence. For the DCS experiment we used 13.09 meV neutrons, which produced an elastic energy resolution of 0.77 meV. For both experiments count rates were normalised using $\rm Y_2Ru_2O_7$ Bragg peaks. This produced absolute measurements of
$\tilde{I}(Q,\hbar\omega)= (\gamma r_0)^2 |\frac{g}{2}F(Q)|^2 2\tilde{\cal S}(Q,\hbar\omega)$
to an overall scale accuracy of 20 \%. Here $\gamma$= -1.913 and $g\approx 2$ are gyromagnetic ratios for the neutron and Ru$^{4+}$ respectively, $r_0$= 2.82 fm is the classical electron radius, and $F(Q)$ is the magnetic form factor. Lacking specific information for Ru$^{4+}$, we used the Ru$^+$ form factor for model comparison.\cite{formfactor}  $\tilde{\cal S}(Q,\hbar\omega)$ is the spherically averaged dynamical correlation function.\cite{lovesey}

\begin{figure}
\includegraphics[width= 8 cm]{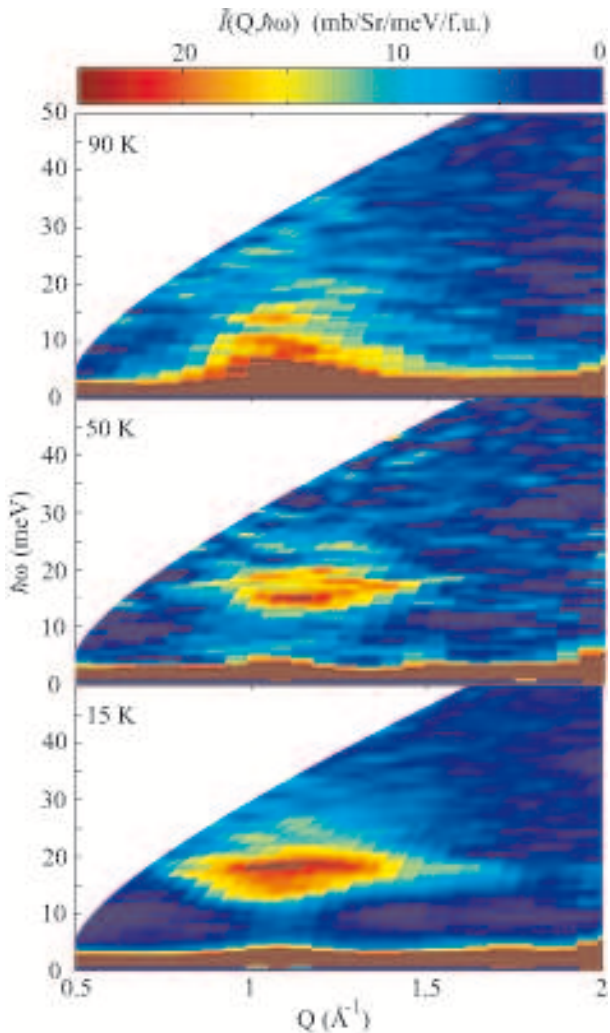}
\caption{Normalized magnetic neutron scattering intensity, $\tilde{I}(Q,\hbar\omega)$, for a Y$_2$Ru$_2$O$_7$ powder sample at $T=$ 90 K (top), 50 K (middle) and 15 K (bottom). Phonon scattering was subtracted as described in the text. The elastic intensity is dominated by nuclear scattering. \label{contour}}
\end{figure}

\begin{figure}
\includegraphics[width= 8 cm]{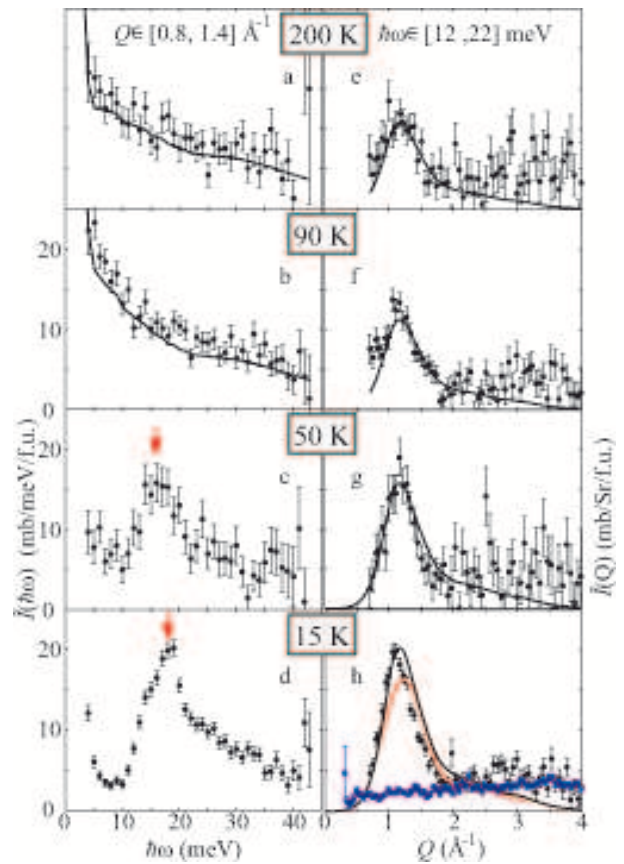}
\caption{$Q$ integrated (a-d) and $\hbar\omega$ integrated (e-h) inelastic neutron scattering intensity of Y$_2$Ru$_2$O$_7$ taken at $T=$ 200 K, 90 K, 50 K and 15 K. Ranges of integration are indicated in each column. The lines show fits to the data described in the text. Lines in (a) and (b) account for the finite sampling of $Q-\omega$ space and partially resolved elastic scattering. Blue squares in (f) are normalized energy integrated data from DCS covering the energy range from 2 meV to 8 meV.\label{cuts}}
\end{figure}

The magnetic neutron scattering cross section for Y$_2$Ru$_2$O$_7$ powder is shown in Fig.~\ref{contour} for $T$= 15 K, 50 K and 90 K. For $T>T_N= 77$~K the spectra indicate short range AFM correlations and a spin relaxation response with a relaxation rate $\Gamma\sim k_BT/\hbar$. Upon cooling below $T_N$, spectral weight moves to higher energies, with modifications up to $4\times k_BT_N$. Fig.~\ref{cuts} shows that the wave vector dependence of the inelastic magnetic scattering is all but unaffected by the comprehensive rearrangement of the excitation spectrum. Indeed the Q-dependence of scattering changes little below room temperature. These data are consistent with frustrated interactions with an energy scale set by $|\Theta_{CW}|= 1100$~K that establish a low energy manifold of near degenerate states with a characteristic inter-atomic ``form factor.''~\cite{ZnCr2O4nature}

The fourier transform of six collinear antiferromagnetically correlated spins on the vertexes of a hexagon provide a reasonable account of the {\bf Q}$-$dependence of inelastic scattering in $\rm ZnCr_2O_4$ over a wide temperature range.\cite{ZnCr2O4_Oleg,ZnCr2O4nature} We have calculated a spherical average of this function for comparison with the powder data in Fig.~\ref{cuts}h. The close correspondence of calculated peak width and position with the data suggest that antiferromagnetically correlated hexagons are also prevalent in $\rm Y_2Ru_2O_7$.

A rigorous understanding of spin correlations in the pyrochlore antiferromagnet is now emerging from theoretical work. For classical spins ($S\rightarrow\infty$), any spin configuration on the pyrochlore lattice where all tetrahedra have zero net moment is a ground state. Higher order spin wave theory ($1/S$ expansion) and the Schwinger-boson large $N$ expansion show that quantum fluctuations favor states that maximize the number of flippable spins. Henley argues that this leads to selection of collinear spin configurations so the low energy physics can be represented by an effective Ising model with a collective spin axis.\cite{henley} The selected spin configurations further are characterized by a zero divergence condition for a polarization field defined on the dual diamond lattice of bonds. The effective Ising model favors AFM hexagons,\cite{effectiveham} which may account for their prevalence in $\rm ZnCr_2O_4$ and $\rm Y_2Ru_2O_7$. The corresponding spin correlation function ${\cal S}({\bf Q})$ is however not the fourier transform of a finite size object as it has infinitely sharp features (pinch points) associated with the divergence free condition.\cite{henley,isakov} Isakov {\em et al} derive an analytic expression for ${\cal S}({\bf Q})$ from the large $N$ approximation, which accurately reproduces results from Monte Carlo simulations of the classical spin Heissenberg model.\cite{isakov}  The black solid lines in Fig.~\ref{cuts} is the spherical average of this expression multiplied by the squared Ru$^{+}$ form factor.\cite{formfactor} The discrepancy at high $Q$ may result from residual phonon contributions to the experimental data and/or the ruthenium magnetic form factor employed in the model calculation. With the overall amplitude as the sole adjustable parameter the large $N$ result provides an excellent account of the data. Note though that the spherical averaging inherent to powder neutron scattering precludes insight on finer details such as the predicted pinch points. At elevated temperatures defects should appear in the form of monopoles\cite{henley} that give rise to a finite divergence of the polarization field and broadening of pinch points\cite{finiteTpinch} which again may not be detectable in powder data. While the theory is not relevant for the ordered state, the close correspondence with the measurements there indicate similarities in local dynamic correlations across the phase transition.

\begin{figure}
\includegraphics[width= 8 cm]{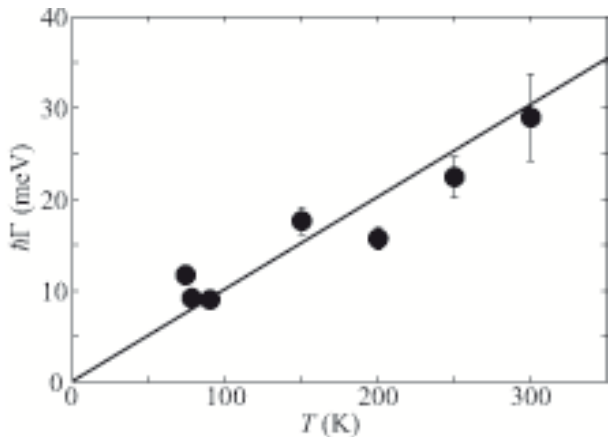}
\caption{$T$ dependence of the spin relaxation rate derived from the global fitting analysis described in the text. The solid line shows the best straight line fit, $\hbar\Gamma= {\cal C}k_BT$ with ${\cal C}= 1.17(9)$.\label{gamma}}
\end{figure}

Although the $Q-$dependence of the data does not change appreciably below room temperatures Fig.~\ref{cuts} shows substantial changes in the fluctuation spectrum. Above $T_N$ the spectra are well described (solid line in Fig.~\ref{cuts}(a-b)) by the following response function:
$\chi''(\omega)=\chi_0\omega\Gamma/(\Gamma^2+\omega^2)$,
which is related to the neutron scattering spectrum through the fluctuation dissipation theorem.\cite{igorandshl} Factorization of the $Q$ and $\omega$ dependence of the cross section was assumed so $\Gamma$ characterizes the local response. The $T-$dependence of the relaxation rate (Fig.~\ref{gamma}) is described by $\hbar\Gamma= {\cal C} k_BT$ where ${\cal C}$= 1.17(9). Linear in $T$ dependence of $\Gamma$ was previously found in simulation of classical Heisenberg spins with AFM interactions on a pyrochlore lattice.~\cite{chalkermoessner} The observation that $k_BT$ is the relevant energy scale indicates a spin system close to quantum criticality and is consistent with the presence of a long length scale from a divergence free polarization field.

For $T<T_N$ a gap appears in the excitaton spectra as spectral weight moves to higher energy (Figs.~\ref{cuts}(c) and ~\ref{cuts}(d)). At $T=15$~K the neutron data indicate an energy gap $\Delta=11(1)$~meV. This is significantly less than the 23 meV gap inferred from specific heat data which may be dominated by activation of the modes associated with the 19.7 meV peak.~\cite{spin_gap} A gapped spectrum was confirmed by the normalized higher resolution DCS data which integrated over energies from 2 meV to 8 meV show no evidence of spin waves emanating from the AFM Bragg peaks (solid blue points in Fig.~\ref{cuts}(h)) as in $\rm ZnCr_2O_4$.\cite{ZnCr2O4prl} When a phase transition breaks a continuous symmetry, Goldstone's theorem assures gapless excitations. The apparent spin gap in the ordered phase thus indicates axial single ion and/or exchange anisotropy. While the gap $\Delta\approx 1.7\times k_BT_N$ surely is an important aspect of the phase transition $\Delta$ is only $\approx$ 12\% of the $k_B\Theta_{CW}$ so spin space anisotropy may be less relevant in the high $T$ phase.
\begin{figure}
\includegraphics[width= 8 cm]{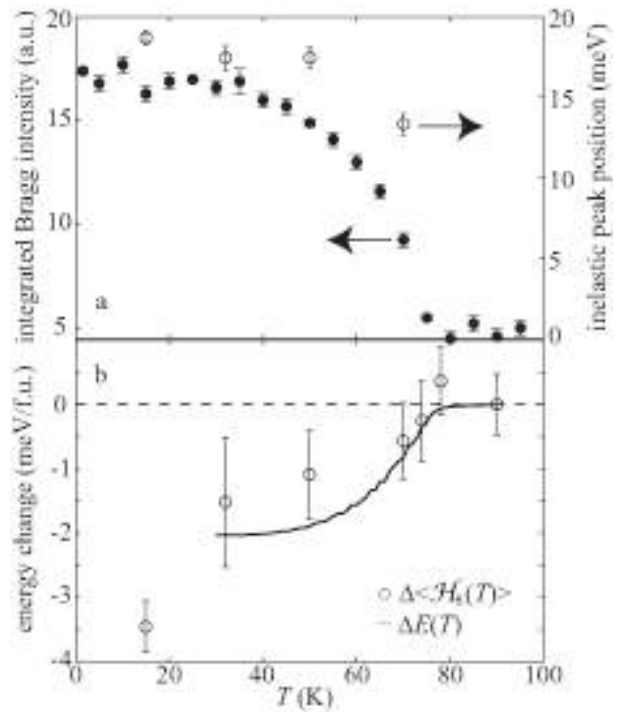}
\caption{(a) $T$ dependence of the integrated intensity of the (111) Bragg peak ($\bullet$) which has nuclear and magnetic contributions. ($\circ$) Indicates the postion of the inelastic peak in the spectrum, derived by fits described in the text. (b) Results of the first moment analysis described in the text. ($\circ$) Indicates the change in spin exchange energy, while the solid line denotes the overall change in energy obtained by integrating specific heat data.\cite{spin_gap} \label{order}}
\end{figure}

Having examined both the high $T$ critical state and the low $T$ ordered state we focus on the transition between these phases. Fig.~\ref{order} compares the $T$ dependence of the inelastic peak position to the integrated intensity of the (111) Bragg reflection. Apart from probing the staggered magnetization squared, the (111) reflection has a small nuclear component and therefore does not go to zero for $T>T_N$.~\cite{diffraction} The inelastic peak position was obtained by fitting a damped harmonic oscillator response\cite{igorandshl} to the $0.8\leq Q\leq 1.4$ \AA$^{-1}$ integrated part of the excitation spectra. While the inelastic peak moves to lower energies upon heating (see Fig. 4) it is important to note that a finite energy peak persists for $T=T_N$. This indicates that short range dynamic correlations at $T_N$ are sufficient to maintain a pseudo-gap in the excitation spectrum.

The shift of spectral weight from a relaxation response towards higher energies implies that the energy of the spin system, $\langle{\cal H}_s\rangle$, is reduced through the phase transition. This can be quantified using normalized inelastic magnetic neutron scattering data and the first moment sum-rule ~\cite{sum-rule,ZnCr2O4prl}
\begin{equation}
\Delta \langle {\cal H}_s\rangle= -\frac{3}{2} \frac{\hbar^2\int_{0}^{\infty} \omega (1-e^{-\beta \hbar \omega})\Delta S(Q,\hbar \omega)\rm{d}\omega}{1-\sin Qd/Qd}.
\end{equation}
Here we consider only the isotropic nearest neighbor interaction between spins separated by a distance $d$. A more complete expression that includes anisotropy terms can readily be derived\cite{igorandshl} but the quality of the present data do not warrant the added complexity. Limiting the integral to 5 to 30 meV and averaging data for $0.8<Q<1.4$ \AA$^{-1}$ the temperature variation of the spin energy $\Delta\langle {\cal H}_s\rangle$ relative to the value at $T=90$~K is shown as open circles in Fig.~\ref{order}b. The data can be compared to the overall change in energy obtained from the magnetic specific heat~\cite{spin_gap}, $\Delta E (T)= \int_0^T C_m(T')dT'-\int_0^{T_{max}} C_m(T')dT'$, shown as a solid line in Fig.~\ref{order}b.  While most of the variation in the thermal energy is accounted for by $\Delta \langle {\cal H}_s\rangle$ and no structural changes have been observed at the transition, the data do leave room for an increase in the lattice energy below $T_N$ by up to 1.5(5) meV/f.u. Note also that spin-phonon coupling was detected through an anomaly in the temperature dependence of a Raman active mode frequency at $T_N$.\cite{bae}

While all known pyrochlore antiferromagnets show some form of (intrinsic or extrinsic) static correlations for sufficiently low $T$, only $\rm Y_2Ru_2O_7$ and its derivatives feature a second order phase transitions to long range N\'{e}el order. The apparent lack of involvement of lattice degrees of freedom indicates that this is a true magnetic instability the understanding of which may require a different frame work than for $\rm ZnCr_2O_4$. The apparent gap in the excitation spectrum indicates that spin space anisotropy may be significant. While we detect the magnitude of the gap there is no direct information about the magnitude or nature of any spin space anisotropic terms in the hamiltonian. The fact that the gap mode is relatively sharp in energy for a powder sample indicates that the corresponding excitations are weakly dispersive and this points to local soft modes pushed to a finite energy on account of quantum fluctuations. Indeed the first moment analysis of $\tilde{\cal S}(Q,\omega)$ shows that the energy gain associated with ordering is intimately related to the development of the 19.7 meV peak. The principal challenge posed by our experiment is thus to understand the interplay between quantum fluctuations and anisotropy terms in establishing the N\'{e}el state spin gap and spin resonance in the frustrated pyrochlore AFM.

We gratefully acknowledge discussions with M. Gingras and O. Tchernyshyov. Work at JHU and ISIS was supported by DoE through DE-FG02-02ER45983 and NSF through DMR-0306940. Work at Rutgers was supported by NSF through DMR-0405682.

\end{document}